
\hoffset -0.35 true in
\hsize = 6.0  true in
\vsize = 8.5 true in
\baselineskip = 14pt plus 2pt   
\parskip = 6pt			
\font\tentworm=cmr10 scaled \magstep2
\font\tentwobf=cmbx10 scaled \magstep2
\font\tenonerm=cmr10 scaled \magstep1
\font\tenonebf=cmbx10 scaled \magstep1
\font\eightrm=cmr8
\font\eightit=cmti8
\font\eightbf=cmbx8
\font\eightsl=cmsl8
\font\sevensy=cmsy7
\font\sevenm=cmmi7
\font\twelverm=cmr12
\font\twelvebf=cmbx12

\def\subsection #1\par{\noindent {\bf #1} \noindent \rm}
\def\mid {\let\rm=\tenonerm \let\bf=\tenonebf \rm \bf}
\def\para{\par \vskip 12 pt}
\def\head{\let\rm=\tentworm \let\bf=\tentwobf \rm \bf}
\def\heading #1 #2\par{\centerline {\head #1} \smallskip
 \centerline {\head #2} \vskip .15 pt \rm}
\def\eight{\let\rm=\eightrm \let\it=\eightit \let\bf=\eightbf
\let\sl=\eightsl \let\sy=\sevensy \let\m=\sevenm \rm}

\def\foots{\noindent \eight \baselineskip=10 true pt \noindent \rm}
\def\sexion{\let\rm=\twelverm \let\bf=\twelvebf \rm \bf}

\def\section #1 #2\par{\vskip 20 pt \noindent {\mid #1} \enspace {\mid #2}
  \para \noindent \rm}

\def\ssection #1 #2\par{\noindent {\mid #1} \enspace {\mid #2}
  \para \noindent \rm}

\def\abstract#1\par{\para \foots {\bf Abstract: \enspace}#1 \para}

\def\author#1\par{\centerline {#1} \vskip 0.1 true in \rm}

\def\abstract#1\par{\noindent {\bf Abstract: }#1 \vskip 0.5 true in \rm}

\def\midsection #1\par{\noindent {\sexion #1} \noindent \rm}

\def\sqr#1#2{{\vcenter{\vbox{\hrule height#2pt
 \hbox {\vrule width#2pt height#1pt \kern#1pt
  \vrule width#2pt}
  \hrule height#2pt}}}}

\def \dk {|\delta_k|}
\def \dm {{\Delta M\over M}}
\def \der {\bigg({\delta\rho\over \rho}\bigg)_H}
\def \dr2 {\bigg({\delta\rho\over \rho}\bigg)_H^2}
\def \HMP {({H_1\over m_{\rm p}})}
\def \hmp2 {\bigg({H_1\over m_{\rm p}}\bigg)^2}

\def \e {\eta }
\def \ee {({\eta\over \eta _0})}
\def \g {\Gamma}

\def \k {({k\e_0 \over 2})}

\def \m {|\mu|}

\def \half {1\over 2}

\def \jetp {Sov. Phys. JETP}
\def \jetpl {JETP Lett.}
\def \sal {Sov. Astron. Lett.}
\def \ap {Astrophys. J.}
\def \apl {Astrophys. J. Lett.}
\def \pd {Phys. Rev. D}
\def \prl {Phys. Rev. Lett.}
\def \pl {Phys. Lett.}
\def \np {Nucl. Phys.}

\def \doublespace {\baselineskip = 20pt plus 7pt \message {double space}}
\def \singlespace {\baselineskip = 13pt plus 3pt \message {single space}}
\singlespace

\def \body {\vfill \eject \parindent = 1.0 true cm
	\ifx \spacing \singlespace \singlespace \else \doublespace \fi}

\def \title#1 {\centerline {{\bf #1}}}
\def \Abstract#1 {\noindent \baselineskip=15pt plus 3pt \parshape=1 40pt310pt
  {\bf Abstract} \ \ #1}





\catcode`@=11
\def \C@ncel#1#2 {\ooalign {$\hfil#1 \mkern2mu/ \hfil $\crcr$#1#2$}}
\def \gf#1 {\mathrel {\mathpalette \c@ncel#1}}	
\def \Gf#1 {\mathrel {\mathpalette \C@ncel#1}}	

\def \gapx {\;\lower 2pt \hbox {$\buildrel > \over {\scriptstyle {\sim}}$}
\; }
\def \lapx {\;\lower 2pt \hbox {$\buildrel < \over {\scriptstyle {\sim}}$}
\; }


\topskip 0.7 true cm

\footline = {\ifnum \pageno = 1 \hfill \else \hfill \number \pageno \hfill \fi}

\def\half{{1\over 2}}
\def\n{\noindent}

\def\v2{\vskip 0.2 true cm}
\def\v3{\vskip 0.3 true cm}

\line{\hfill \foots{\bf IUCAA Preprint -27/92}}

\vskip .45 true in

\heading {Post-COBE predictions for Inflationary }

\heading { Gravity Wave and  Density Perturbation spectra
\footnote*{\twelvebf{To
appear in the Proceedings of the First Iberian Meeting on
Gravity, September 21 - 27, 1992, Evora, Portugal.
Ed. M. Bento, O. Bertolami and J. Mourao (World Scientific
Publishing Co.).}} }

\vfill
\smallskip
\centerline{{\bf VARUN SAHNI} $^{\dag}$ and  {\bf TARUN SOURADEEP} $^*$}
\medskip
\centerline{Inter-University Centre for Astronomy and Astrophysics}
\centerline{Post Bag 4, Ganeshkhind, Pune 411 007}
\centerline{INDIA}
\vfill
\vskip .25 true in
\centerline {\bf ABSTRACT}
\bigskip

We assess the relative contribution to the COBE - measured microwave
anisotropy arising both from relic gravity waves as well as primordial
density perturbations originating during inflation.
We show that the gravity wave contribution to the CMBR anisotropy
depends sensitively upon $n$ -- the primordial spectral index
($\dk^2 \propto k^n$), increasing as $n$ deviates from a Harrison - Zeldovich
spectrum (n = 1). As a result, for
$n < 0.84$ the contribution from gravity waves towards  $\delta T/ T $
is greater than the corresponding contribution from density
perturbations, whereas for $n > 0.84$ the reverse is true.
($ n = 0.84$ corresponds to an expansion index $p = 13.5$ in models with
power-law inflation $a \propto t^p$. ) Our results show that for
a scale-invariant Harrison-Zeldovich spectrum generated by chaotic
inflation, gravity waves contribute approximately $24\%$ to the CMBR anisotropy
 measured by COBE. Applying our results
to the cold dark matter scenario for galaxy formation, we find that
in general CDM models with tilted power spectra ($n < 1$), require the
biasing parameter to be greater than unity, on scales of $16h_{50}^{-1}
Mpc$.
We also obtain an expression for the COBE - normalised amplitude
and spectrum of the stochastic gravity wave background and compare it
with the sensitivity of planned laser-interferometer gravity wave
detectors.


\vskip 1.0 cm

\vskip 0.25cm

\settabs 10\columns
\+&&&email~:&$\dag$~~~~varun@iucaa.ernet.in \cr
\+&&&&$^*$~~~~tarun@iucaa.ernet.in\cr
\medskip

\vfill\eject
\topskip  0.7cm
\bigskip

\vfil\eject

\nopagenumbers
Both gravity waves and adiabatic density perturbations,
characterised by a power spectrum $\dk \propto k^n$, are generically predicted
by the Inflationary Universe scenario. On entering our horizon during the epoch
of matter-radiation decoupling, density perturbations as well as gravity waves
cause
distortions in the cosmic microwave background radiation (CMBR) due to the
well known Sachs-Wolfe effect. The  anisotropy in the CMBR recently observed
 by COBE measures the amplitude of gravity waves as well as density
perturbations
 providing thereby a direct probe of the inflationary epoch of our universe.

The close similarity between gravity waves
and massless scalars, in a Friedman - Robertson - Walker (FRW) background
originally pointed out by Grishchuk$^{1}$,
makes it possible to express each polarisation state of the graviton
in terms of solutions to the massless, minimally coupled Klein-Gordon
equation: $h_{\times(+)} = \sqrt{8\pi G}~\phi_k\,{\rm e}^{-i{ \bf k}.{\bf
x}}\,$

$$
\ddot \chi_k + [k^2 - {\ddot a\over a}]\, \chi_k = 0
\eqno(1a)
$$

\noindent where $\chi_k = \phi_k \times a(\e)$,
$a(\e)$ is the scale factor of the Universe:
$ ds^2 = a^2 (\e)(d\e^2 - d{\bf x }\, ^2)$,
and $k$ is the comoving wavenumber $k = 2\pi a/\lambda$.
Eq.(1a) closely resembles the Schroedinger equation in quantum mechanics,
with ${\ddot a / a}$, playing the role of the potential barrier `V'.
(The form of $V \equiv (\ddot a/ a)$, is shown in Figure 1, for inflation
followed by a matter dominated epoch.) Solutions of equation (1a) have some
interesting properties, for instance, small wavelength solutions of (1a)
are adiabatically damped:

$$\phi_k ^ + (\e) \equiv {\chi_k\over a(\e)} ~~{\mathop{=}_{k\e \gg 2\pi
} } {1 \over \sqrt{ 2k} a(\e) }~{\exp (-ik\e)}, \eqno(1b)$$
whereas long wavelength modes asymptotically approach the form :
$$
\phi_k (\e) \equiv {\chi_k\over a(\e)} ~~ {\mathrel{\mathop{=}_{k\e \ll 2\pi} }
}
 A(k) + B(k) \int {d\ee \over a^2}~,
 \eqno(1c)
 $$
 \noindent where A and B are constants. The form of (1c) implies
that the amplitude of a fluctuation
 (equivalently -- gravity wave), freezes to a constant value when its
 wavelength (during inflation), becomes larger than the corresponding
 Hubble radius. Since modes are continuously being pushed outside the
 Hubble radius during inflation, a mode with a fixed wavenumber ``k", which
left
 the Hubble radius during inflation ($k = 2 \pi ~ \eta_1^{-1}$),
 will re-enter it later
 during the epoch of matter domination ($k = 2 \pi ~\eta_0^{-1}$).
 The amplitude
 of the mode between these two epochs is super-adiabatically amplified,
 which in field-theoretic language corresponds to the quantum creation of
 gravitons$^{1}$ (see fig. 1).

 The possibility that relic gravity waves
 may be generated during inflation, was first investigated by
 Starobinsky$^{2}$, who showed that the spectrum of gravitons is scale
 invariant in a fairly broad wavelength interval, if inflation proceeds
 exponentially and is followed by a radiation dominated epoch.
 The quantum creation of relic gravity waves was subsequently
 extended to other inflationary models by
 Abbott \& Wise$^{4}$ and
 Sahni$^{6}$. (The effect of a post inflationary matter dominated epoch
 on the graviton spectrum was studied by Allen$^{5}$, and also by
  Sahni$^{6}$.)
 In particular it was demonstrated that,
 for power law expansion $a = (t/t_0)^p \equiv
 ({\e\over \e_0})^{1-2 \nu\over 2},\phantom{.}
 ( \nu - {3\over 2} = {1 \over {p - 1}}  = - 3{{1 + w}\over {1 + 3w}} , $
 where $w = {P\over \epsilon}$ is the equation of state of matter)
 \phantom{.}  Eq.(1a) can be solved exactly, with the following solution
 describing positive frequency waves in an FRW Universe in the adiabatic
 limit:
 $$
 \phi_k ^ + (\e) = \sqrt {{\pi \e_0\over 4}}\, \left({\e\over
 \e_0}\right)^\nu\,
 H_\nu ^{(2)} (k\e) \,
 \eqno(1d)
 $$

($\nu = {3\over 2}$ corresponds to exponential inflation, $w = - 1$;
$\nu > {3\over 2}$ to power law inflation, $- 1 < w < -{1\over 3}$;
${1\over 2} < \nu < {3\over 2}$ to
super-exponential inflation (pole - driven inflation), $ w < -1$; and
$ -{3\over2} \le \nu \le 0$ to
matter dominated expansion, $ 0 \le w \le 1$ .)
{}From (1c) and (1d) it can be shown that, the amplitude of gravity waves
on scales greater than
the Hubble radius, has the time independent form$^{6}$

\vfil\eject

$$\eqalign{  h_{+,\times}  = &\sqrt{16 \pi G}
{k^{3 \over 2}\over \sqrt 2 \pi} |\phi_k|\Big|_{k|\e _1| = 2\pi}
= \sqrt{16 \pi} A(\nu) ~\bigg({H_1\over m_{\rm p}}\bigg),\cr & A(\nu) =
\g (\nu)\pi^{-\nu}(\nu - \half)^{-1},\cr}
\eqno(2)
$$

\noindent where we have summed over both polarisation states of the graviton.
The value of $\nu$ is  related to the spectral index $n$ of density
fluctuations
as $\nu = (4 - n ) /2 $ and
$H_1$ is the value of the Hubble parameter at a time $k|\e _1| = 2\pi$,
when the given mode left the Hubble radius.
For exponential inflation
$\nu = {3\over 2}$ and $h_{+,\times} = (2/\sqrt{\pi}) ( H_1/m_{\rm p}) $.
 ~($m_{\rm p}$ is the planck mass, $m_{\rm p} = 1.2 \times 10^{19} GeV$.)

Both long wavelength gravity waves as well as adiabatic density
perturbations generated during inflation, create potential fluctuations
at the surface of last scattering, when their wavelength becomes smaller
than the horizon size during the matter dominated epoch.
This leads to distortions in the cosmic microwave background, described
by the well known Sachs - Wolfe effect
$^{4, 7}$ $(\delta T/ T) \sim {1\over 3}~ (\delta \varphi /c^2)$. Writing
$(\delta T /T)$ in terms of a multipole expansion:

$${\delta T \over T} = \sum_{l,m}{\rm a}_{lm} Y_l^m(\alpha,\delta), \eqno(3)$$

\noindent it can be shown that, for gravity waves$^{4,8}$

$$ {\rm a}_2^2 \equiv \langle|{\rm a}_{2m}|^2\rangle = 0.145~
h_{+,\times}^2 ~ g(n), ~~~g(n) = 5770.3 \times \int_0^\infty db~
\bigg({b \over 2}\bigg)^{n - 2} [I_2(b)]^2  ,\eqno(4a)$$

\noindent where$^{8}$

$$ I_l (b) = \int^b_{k \eta_{rec}} {dy \over (b - y)}~
 { J_{l +{1 \over 2}} (b -y)  \over (b - y)^{3 \over 2} }
 {J_{{5 \over 2}} (y) \over y^{3 \over 2} },~~~~ b = k\eta_0. \eqno(4b)$$

\noindent We have numerically integrated (4b) to obtain $g(n)$
($k\eta_{rec} \approx 0$),
which is well approximated by the fitting formula
$ g(n) = {\rm exp} [0.6(n - 1)]$
over the range $0.5 \le n \le 1$.
\noindent The corresponding value of the {\it rms} quadrupole amplitude is
$$
Q_{T}^2 = {5\over 4 \pi} ~{\rm a}_2^2~ F_2 = 2.9~ A^2(\nu)~
\bigg({H_1\over m_{\rm p}}\bigg)^2 g(n)~F_2~.\eqno(4c)$$

\noindent (For exponential inflation $\nu = {3\over 2}$, and
 $Q_{T}^2 = (2.9 / 4 \pi^2) (H_1/ m_{\rm p})^2~F_2.$)
(The subscript ``T", denotes the fact that the quadrupole anisotropy is induced
by {\it tensor} waves, gravity waves being quadrupolar, do not contribute to a
dipole component ${\rm a}_1$.) $F_2$ incorporates the finite beam width of the
COBE-DMR instrument ($F_2 \approx 0.99$) .

Scalar density perturbations generated during inflation,
 are characterised by a spectrum $\dk^2 = {\cal A}~ k^n$, the scale
 invariant Harrison - Zeldovich spectrum $n = 1$, is predicted by inflationary
models
with exponential expansion.  Models with power law inflation
($a(t) \propto t^p, p > 1$), which arise in a number of theories
including extended inflation$^{9}$, predict a more
general value for the spectral index$^{10}$ , $ n = (p - 3)/ (p - 1) <
1~~~$ . ($n > 1$ , is predicted for super-exponential inflation$^{28}$.)

The amplitude of density perturbations at horizon crossing ($k = H_0/c$)
is given by$^{11}$
$$
\dr2 \equiv {k^3 \dk^2 \over 2 \pi^2} = { {\cal A} \over 2 \pi^2} H_0^{n+3}.
\eqno(5)
$$
(we assume $c = 1$, for simplicity.)
The associated fluctuations in the microwave background are given by  ( for $l
\ge 2$)
$$
{\rm a}_l^2 \equiv \langle|{\rm a}_{lm}|^2\rangle = {H_0^4\over 2\pi}
\int_0^\infty {dk\over k^2}\dk^2 j_l^2(kx)
\eqno(6a)
$$
where $x = (2/ H_0)$, is the present day horizon size
($j_l^2(kx)$ are spherical Bessel functions). For power law
spectra $\dk^2 = {\cal A}~k^n$, we obtain$^{12}$
$$
{\rm a}_2^2 = { {\cal A} H_0^{n+3}\over 16}~f(n) = {\pi^2\over 8} f(n) \dr2 ,
{}~~~ f(n) = {\Gamma(3 - n) \Gamma ({3+n\over 2})\over
\Gamma^2({4-n\over 2}) \Gamma({9-n\over 2})}.
\eqno(6b)
$$

\noindent Values of $n$ in the range $0.5 \le n \le 1.7$, are consistent with
the recent COBE results$^{13}$.
For $n = 1$, (6b) reduces to
${\rm a}_2^2 = (\pi/12)~(\delta\rho/\rho)^2_H $.

\noindent The related value of the {\it rms} quadrupole amplitude is given by
$$
Q_{S}^2 = {5\over 4\pi} ~{\rm a}_2^2 ~F_2 = {5\pi \over 32} f(n) \dr2 ~ F_2
\eqno(7)$$

\noindent (the subscript ``S" refers to {\it scalar} perturbations.)

As shown by a number of authors, scalar field fluctuations which left the
Hubble radius during inflation, upon re-entering the horizon during
radiation (matter) domination, give rise to a density fluctuation, whose
{\it rms} value is given by$^{14}$
$$
\der = b~ {H_1\over \dot \phi} ~\delta \phi
\eqno(8)
$$
where $H_1$ is the Hubble parameter at a time $t_1$, when a scale which
reenters the horizon at $t_0$, left the Hubble radius during the inflationary
epoch (see Figure 1).  For modes entering the horizon during radiation (matter)
 domination, $b = 4 ({2\over 5})$. (The main contribution to the observed
quadrupole anisotropy however, comes from modes entering the horizon during
matter domination, we shall therefore assume $b = {2\over 5}$ in the
ensuing discussion.) $\delta \phi$ arises because of quantum fluctuations
in the scalar field during inflation. For modes entering the horizon today
$\delta \phi$ is given by $\delta \phi = A(\nu) H_1$ (see (2)), from which we
recover the standard result $\delta \phi = {H_1 \over 2\pi}$, for exponential
inflation$^{14, 15}$.

For power law inflation, the inflaton potential has the form
$ V(\phi) =  \break V_0 \exp(- \sqrt{16\pi/ p}~(\phi/m_{\rm p}))$
which results in the following exact solution$^{16}$ to the field
equations: $H = (p/t)$, $(H/\dot\phi) = (\sqrt{4\pi p}/ m_{\rm p})$,
as a result

$$
\der =   A(\nu)~{\sqrt{16\pi p}\over 5}~ {H_1\over m_{\rm p}}.
\eqno(9)
$$

\n Combining  Eqs. (7) and  (9), the {\it rms} quadrupole amplitude  of
the  temperature anisotropy
$$
Q_{S}^2 = {\pi^2 \over 10}~p~  A^2(\nu) \hmp2 f(n)~F_2
\eqno(10)
$$
$(\nu - {3\over 2} =  (p - 1)^{-1} = (1 - n)/ 2)~), $
for power law inflation. (See also Fabbri, Lucchin and Mataresse;
Lyth \& Stewart.)$^{17}$
The above result for power law inflation can  be extended to cover models
of slow-roll (quasi-exponential)  inflation . As shown in
 Souradeep and Sahni $^{18}$,  $Q_S^2$ for
a chaotic inflation model with a $m^2 \phi^2$ potential corresponds to  a power
law model with $p = 2 N$, where $N = H \Delta t$ is the number of e-foldings
($\Delta t$ is the duration of the inflationary epoch and $H \approx constant$
 is the Hubble parameter during inflation). In terms of the number of
e-foldings in
a model of slow-roll inflation, the following relation

$$ 2 N = p = {n -3 \over n -1 },\eqno(11)$$

\n fixes the spectral index of density perturbations  $n$. As a result
 slow-roll inflation ( chaotic $m^2 \phi^2$ model) with minimal e-foldings
$N = 67$ is equivalent to a power law model with $p = 134$ which corresponds to
a spectral index $n = 0.985$.

Since the coefficients ${\rm a}_{lm}$ in (3) are Gaussian random variables,
the
variances $Q_{T}^2$ and $Q_{S}^2$ given by (4c) and (10) must be added,
when considering the combined effect of both gravity waves and adiabatic
density perturbations on the anisotropy of the CMBR.
As a result, the net quadrupole anisotropy will be given by
$$
Q_{\rm COBE-DMR}^2 = Q_{S}^2 + Q_{T}^2
\eqno(12)
$$
where $Q_{\rm COBE-DMR}$ is the {\it rms}-quadrupole normalised amplitude
$Q_{rms-PS} = 5.86\times 10^{-6}$ detected by COBE $^{13}$.
{}From (4c) and (10), we find that the ratio   $(Q_{T}^2/
Q_{S}^2) = (2.94~ (n - 1) g(n) /(n -3) f(n))$ depends sensitively on $n$
-- the spectral index of density perturbations. We find that for $n \ge 0.84$,
the
contribution from gravity waves to the CMBR
anisotropy dominates the contribution from scalar density perturbations,
whereas for $n < 0.84$ the reverse is true. ($n = 0.84$ corresponds to $p =
13.5$.)
The ratio $(Q_{T}^2 / Q_{S}^2)$ has been plotted against  $n$
in Figure~2 . For slow-roll (chaotic) inflation we find that the microwave
distortion
due to gravity waves amounts to  $\simeq 31\%$ of the distortion
caused by density fluctuations, in agreement with earlier work by
Starobinsky $^8$ and contrary to recent claims by Krauss and
White$^{19}$.

Substituting the values of
$Q_{T}^2$ and $Q_{S}^2$ in Eq.(12), we obtain
$$
\bigg({H_1\over m_{\rm p}}\bigg)^2 = \bigg[{\pi^2 \over 10} ~ {n - 3 \over n -
1}~f(n) +
 2.9 ~g(n) \bigg]^{-1} [A(\nu)]^{-2} ~{Q_{\rm COBE-DMR}^2 \over F_2}
\eqno(13)
$$
for power law inflation. Setting $p=2 N$ in  Eq. (13) we can recover
 the result for slow-roll  inflation$^{18}$.
As a result Eq.(13) allows us to determine the Hubble parameter during
inflation both for power law, as well as for quasi-exponential inflation.

\topskip 6.0  true in

The value of $( H_1/m_{\rm p})$ allows us to determine the value of
$Q_S^2$ -- the fraction of the quadrupole anisotropy contributed
 by density perturbations, as well as $(\delta\rho/\rho)_H^2$ -- the
 amplitude of density  fluctuations at horizon crossing (see (5), (9)).
The value of  $(\delta\rho/\rho)_H$ fixes the normalisation ${\cal A}$ of the
power
spectrum of density fluctuations to be

$$ { {\cal A} \over 2 \pi^2} = \bigg[ {72.5 \over 16\pi} ~ {n - 1 \over n - 3}~
{}~g(n)
 + {5\pi \over 32}
{}~f(n)\bigg]^{-1} \bigg( {c \over H_o }\bigg)^{n + 3} { Q_{\rm COBE-DMR}^2
\over F_2}
 ~,\eqno(14)$$

\noindent ( we have reintroduced $c$ in the above  expression for completeness
 and as noted earlier (Eq. 4c), the finite beam width factor $F_2 \approx 0.99$
 for COBE-DMR ).
We now use the COBE-DMR results to normalise the spectrum and
 predict the value of the {\it rms} mass fluctuation on a given
 scale

$$\bigg({\Delta M\over M}\bigg)^2(R) = {1\over 2\pi^2} \int{k^2 dk
  \dk^2 W^2(kR)},
 \eqno(15)$$

 \noindent where $W(kR)$ is the {\it top hat} window function
 $^{20}$.

\topskip 0.7 true  cm

Our results for cold dark matter models with power law
primordial spectra :
$\dk^2 = {\cal A}\, k^n~ T_{k,{\rm cdm}}$, (where $T_{k,{\rm cdm}}$ is
the CDM transfer function given in Appendix~G of Bardeen et
al.$^{21}$~)
are shown in Figure  3, for different values
of the parameters $n$ and $h_{50}$ ($\Omega_{\rm cdm} \simeq
0.9, \Omega_{\rm baryonic} \simeq 0.1$, is assumed).
(Values of n in the range $0.5 \le n \le 1.7$, are consistent
with the COBE - DMR results $^{13}$.)
For a scale-invariant spectrum $n = 1$,
our results agree with those of Bond and Efstathiou$^{22}$.
Our results also show that $\dm (16h_{50}^{-1} Mpc) \le 1$ for
$n \le 0.87$. Since mass need not trace light, in models with
nonbaryonic
dark matter, our results can be taken to mean that a biasing
factor
($b_{16} = (\dm)^{-1}_{16}$) mostly
greater than unity, is required in order to reconcile
theoretical models
based on power law inflation, with observations. From Figure  3 it
follows
that the value of $b_{16}$ is sensitive to both $n$ as well as $h_{50}$,
decreasing with increasing $n$ and $h_{50}$. It may be noted that since
 gravity waves contribute predominantly to the
 CMBR anisotropy for $n \le 0.84$, the biasing factor
 $b_{16h_{50}^{-1}}$
 is much larger than it would have been had only the contribution from
 scalar density perturbations to the CMBR been considered. This makes
 CDM
 models with power law inflation less compatible with the excess galaxy
 clustering observed in the APM survey, than had previously been assumed
 $^{23}$.

Having normalised the amplitude of scalar density perturbations
generated during inflation, we now proceed to evaluate the normalised
spectrum of relic gravity waves. During inflation ({\it ie} for
 $\e < \e_0 < 0$), gravitons are described
by Eq.(1d) which represents the state corresponding to the adiabatic vacuum .
During the matter dominated epoch however, gravitons
are described by a state which is
a linear superposition of positive and negative frequency solutions of Eq.(1a):

\vfil\eject

$$
\tilde \phi_k (\e) = \alpha \tilde \phi_k^{(+)} + \beta \tilde \phi_k^{(-)}
\eqno(16a)
$$
where
$$
\tilde \phi_k ^ {(+,-)} (\e) = \sqrt {{\pi \e_0\over 4}}\,
\left({\e\over \e_0}\right)^\mu\,
H_{\vert \mu \vert} ^{(2,1)} (k\e) \, \eqno(16b)
$$
$\e > \vert \e_0 \vert $. $\mu$ can be
related to the equation of state of matter $ \mu =  {3\over 2}\, {{w -
1}\over {1 + 3w}}$ where $ w = P/\epsilon$,
($-{3\over 2} \le \mu \le 0$ for reasonable equations of state for
matter: $ 0 \le w \le 1 $).

The Bogoliubov coefficients $\alpha$ and $\beta$ can be obtained by
matching (1d) and (16) at wavelengths larger than the horizon size via
(1c).
As a result we obtain for $k\e_0 < 2\pi$,

$$
\alpha \pm \beta =   \pm i\gamma^{\pm 1} \phantom{.}\k^{\mp(\nu + |\mu|)}
\eqno(17)
$$
where $\gamma = \pi^{-1}\g (\nu)\,\g (1 + |\mu|),$ and $|\alpha|^2 -
|\beta|^2 = 1$, (see Sahni$^6$ (1990) for details).
(For $k\e_0 > 2\pi$ the adiabatic theorem gives  $\alpha \simeq 1,
\beta \simeq 0$.)

The energy density of created gravitons can now be determined exactly by
substituting the expression for $\tilde \phi_k (\e)$ in (16), with the
values of the  Bogoliubov coefficients given in Eq.(17) in the
expression$^{29}$
$$
\epsilon_g = \langle T^0_0\rangle = {1\over 2 \pi^2 a^2}
\int dk\,k^2\left(\left\vert\dot
{\tilde\phi}_k
\right\vert^2 + k^2 |\tilde\phi_k|^2 \right) ,
\eqno(18)
$$
(both polarisation states of the graviton have been included in Eq.(18).)

Of greater relevence to us is the spectral energy density of gravity
waves, $\epsilon(\omega) = \omega{d\epsilon_g\over d\omega}$ which
can be derived from Eq.(18) and has
the simple form$^6$
$$
\epsilon(\tilde\omega) = b_2^2\,{\tilde\omega}^{1 -
2\tilde\nu}\HMP^2\epsilon_m
\quad \hbox{     for  } 1 < \tilde\omega < {3\over
4\pi}\Omega_r^{-\half} ,
$$
$$
\epsilon(\tilde\omega) = b_1^2\,{\tilde\omega}^{3 -
2\tilde\nu}\,\HMP^2\epsilon_r
\quad \hbox{  for    }
{3\over 4\pi}\,\Omega_r^{-\half} < \tilde\omega
\eqno(19)
$$
where $\tilde\omega = {k\e\over 2\pi} = {\lambda_h\over \lambda},$
is the dimensionless wavenumber expressed in units of the horizon scale
($\lambda$ being the physical wavelength and $\lambda_h$ being the
present scale of the horizon, $\lambda_h \simeq 2.\,10^{28}
h_{50}^{-1}\,cm$.).  $1 < \tilde\omega < {3\over
4\pi}\Omega_r^{-\half}$ corresponds to wavelengths larger than the
horizon size at matter radiation equality,
$\epsilon_m$ and $\epsilon_r$ are the energy densities of matter and
radiation respectively; $\epsilon_m \approx \epsilon_{cr} = 4.2\,.
10^{-9} h_{50}^2$ ergs/cm$^3$,
$\Omega_r = {\epsilon_r\over \epsilon_m} \simeq 10^{-4}\,h_{50}^{-2}$,
$h_{50}$ is the present value of the Hubble parameter in units of
50 km/sec/Mpc.

\topskip 6.5 true in

{}From Eq.(19) we see that the spectrum of the gravity wave background is
uniquely
specified once
$\HMP$ -- the dimensionless value of the Hubble parameter during
inflation is established from Eq.(13).
The ratio of the spectral energy density of gravity waves to the
critical energy density
$~\Omega_g(\lambda) = \epsilon(\lambda)/ \epsilon_{cr}~~$ and the
dimensionless amplitude of gravity waves, $h_{\times +} (\lambda)$
 have been plotted
in Figure $4$, for gravity waves created during exponential inflation,
slow-roll (quasi-exponential) inflation and power law inflation.
(For the corresponding analytical expressions describing $\Omega_g(\lambda)$,
 see Sahni$^{6}$ and Grishchuk \& Solokhin $^{24}$.)
 The quantities $h_{\times +} (\lambda)$ and  $\Omega_g(\lambda)$
are directly related by  $ h^2_{\times +} (\lambda)  =
  24 (\lambda / \lambda_h) \Omega_g(\lambda) $.
We find that the amplitude of gravity waves
is significantly smaller than the sensitivity
of the current generation of terrestrial bar and beam detectors $^{25}$.
The best hope for the detection of the stochastic gravity wave background
appears to lie with the {\it space - interferometer}, whose development
is still in the conceptual stage $^{26}$.

After completing this work, we became aware of papers$^{27}$ by
Davis et al.; Salopek; Liddle \& Lyth;
Lucchin, Mataresse
\& Mollerach ; and Lidsey and Coles,
in which results overlapping with those of this paper and
ref [18] were obtained.

\vfil\eject

\topskip 0.7 true cm

\vskip .4cm

\vfill \eject
\section  { References}

\vskip .4cm
\n
\item{1.} Grishchuk, L.P., {\it Zh. Eksp. Teor. Fiz.} {\bf 67}  (1974) 825
 [~{\it \jetp} {\bf 40,}(1975) 409~]
\vskip .2 cm

\item{2.} Starobinsky, A.A., {\it \jetpl} {\bf 30}  (1979)  719
\vskip .2 cm

\item{3.} Sahni, V.,  {\it Class. Quantum Grav.} {\bf 5}  (1988) L113
\vskip .2 cm

\item{4.}Abbott, L.F. and Wise, M.B.,  {\it \np} {\bf 135B} (1984) 279
\vskip .2 cm

\item{5.}Allen, B.,  {\it \pd} {\bf 37}  (1988) 2078
\vskip .2 cm

\item{6.} Sahni, V., {\it \pd} {\bf  42}  (1990) 453
\vskip .2 cm

\item{7.} Sachs, R.K. and Wolfe, A.M.,  {\it \ap} {\bf 147}  (1967) 73
\vskip .2 cm

\item{} Rubakov, V.A., Sazhin, M.V. and Veryaskin, A.V., {\it \pl} {\bf 115B}
(1982) 189
\vskip .2 cm

\item{}Fabbri, R. and Pollock, M.D., {\it \pl} {\bf 125B} (1983)  445
\vskip .2 cm

\item{8.}  Starobinsky, A.A., {\it Pis'ma Astron. Zh.} {\bf 11} (1985) 323
[~{\it \sal}
{\bf 11} (1985) 133 ~]

\item{9.} La, D. and Steinhardt, P.J.,  {\it \prl} {\bf 62}  (1989) 376
\vskip .2 cm

\item{10.} Lucchin, L.F. and Mataresse, S., {\it \pl} {\bf 164B}  (1985) 282
\vskip .2 cm

\item{} Lucchin, L.F. and Mataresse, S., {\it \pd} {\bf 32}  (1985) 1316
\vskip .2 cm

\item{11.} Kolb, E.W. and Turner, M.S., {\it The Early Universe},
(Addison - Wesley Publishing Company 1990)
\vskip .2 cm

\item{12.}Gradshteyn, I.S. and Ryzhik, M., {\it Table of Integrals, Series and
Products}, (Acad. Press Inc., N.Y., 1980) p. 692
\vskip .2 cm

\item{13.} Smoot, C. F.  et al., {\it Astrophys. J. Lett.} {\bf 396}  (1992) L1
\vskip .2 cm

\item{14.} Starobinsky, A.A.,  {\it \pl} {\bf 117B}  (1982) 175
\vskip .2 cm

\item{} Guth, A. and Pi, S.Y., {\it \prl} {\bf 49}  (1982) 1110
\vskip .2 cm

\item{} Hawking, S.W., {\it \pl} {\bf 115B} (1982) 295
\vskip .2 cm

\item{} Bardeen, J.M., Steinhardt, P., and Turner, M., {\it \pd} {\bf 28}
(1983)  679
\vskip .2 cm

\item{15.}Bunch, T. S. and Davies, P. C. W., {\it Proc. R. Soc.} {\bf A360}
(1978) 117
\vskip .2 cm

\item{} Linde, A.D., {\it Phys. Lett.} {\bf 116B} (1982)  335
\vskip .2 cm

\item{} Vilenkin, A. and Ford, L. H., {\it Phys.  Rev.} {\bf D25}  (1982) 1231
\vskip .2 cm

\item{16.}Burd, A.B. and Barrow, J.D., {\it \np} {\bf B308} (1988) 929
\vskip .2 cm

\item{17.}Fabbri, R., Lucchin, L.F. and Mataresse, S., {\it \pl} {\bf 166B}
(1986) 49
\vskip .2 cm

\item{} Lyth, D.H. and Stewart, E.D., {\it \pl} {\bf 274B} (1992) 168
\vskip .2 cm

\item{18.} Souradeep, T. and Sahni, V.,  IUCAA preprint/July-07-92 (1992),
\vskip .2 cm
\item{}Souradeep, T. and Sahni, V., {\it Mod. Phys. Lett.} {\bf A7} (1992)
3541
\vskip .2 cm

\item{19.} Krauss, L. M. and White, M., {\it \prl} {\bf 69} (1992) 869
\vskip .2 cm

\item{20.} Peebles, P.J.E., {\it The Large Scale Structure of the Universe}
(Princeton University, Princeton 1980)
\vskip .2 cm

\item{21.}Bardeen, J.M., Bond, J.R., Kaiser, N, and Szalay, A.S.,{\it \ap}
{\bf 30} (1986) 15
\vskip .2 cm

\item{22.}Bond, J.R. and Efstathiou, G.,  {\it \ap} {\bf 285} L45, (1984) L45
\vskip .2 cm

\item{}Bond, J.R. and Efstathiou, G., {\it MNRAS} {\bf 226}  (1987) 655
\vskip .2 cm

\item{23.} Liddle,  A. R., Lyth, D. H., and Sutherland W. J. , {\it Phys.
Lett.}
 {\bf B274} (1992) 244
\vskip .2 cm

\item{24.} Grishchuk, L.P. and Solokhin, M., {\it \pd} {\bf 43} (1991) 2566
\vskip .2 cm

\item{25.} Christensen, N., {\it \pd} {\bf 46} (1992) 5250
\vskip .2 cm

\item{26.} Thorne, K.S., in {\it 300 Years of Gravitation}, eds. S.W. Hawking
and W. Israel, (Cambridge Univ. Press, Cambridge, 1988) p.330

\item{27.} Davis R.L., Hodges, H.M., Smoot, G.F., Steinhardt, P.J. and
Turner, M.S., {\it \prl} {\bf 69} (1992) 1856
\vskip .2 cm

\item{} Salopek, D. S., {\it \prl}  {\bf 69} (1992) 3602
\vskip .2 cm

\item{} Liddle, A. R. and Lyth, D. H., {\it \pl} {\bf B291} (1992) 391

\item{} Lucchin, F., Mataresse, S. and Mollerach, S.,  {\it \apl} {\bf 401}
(1992) L49
\vskip .2 cm
\item{} Lidsey, J.E. and Coles, P., {\it MNRAS} {\bf 258} (1992) 57p
\vskip .2 cm

\item{28.} Gasperini, M. and Giovannini, M., {\it \pl} {\bf B282} (1992)  36
\vskip .2 cm
\item{} Gasperini, M. and Veneziano, G., CERN preprint: CERN-TH.6572/92

\item{29.} Ford, L.H. and Parker, L., {\it \pd} {\bf 16} (1977) 1601
\vskip .2 cm

\section { Figure Captions}

\leftline {\bf Fig. 1}
\vskip .2 cm
The superadiabatic amplification of gravity waves is shown for a ``potential
barrier" $V[a(\eta)] = (\ddot a/ a)$, for inflation $~(a = (\e_o/\e)~)$
followed
by radiation and matter domination$~^{3}$ $~~(a = a_o \e ( \e + \tilde
\e_o)~)$.
At early times $t < t_1$, the scalar field is in its vacuum state $\tilde
\phi_k^{(+)}$.
At late times $t \approx t_0 $, the scalar field will in general be described
by a
linear superposition of positive and negative frequency solutions of eq. (1) :
$\tilde \phi_k (\e) = \alpha \tilde \phi_k^{(+)} + \beta \tilde \phi_k^{(-)}.$
A mode with comoving wavenumber $k = (2\pi a/\lambda)$ is  shown to leave
the Hubble radius during inflation ($t_1$),
and re-enter it during matter domination ($t_0$). (Figure not drawn to scale.)

\bigskip

\leftline {\bf Fig. 2}
\vskip .2 cm

The ratio of the Quadrupole anisotropy from tensor
 and scalar waves is plotted against both spectral index $n$ and   $p$~
(in $a \propto t^p$) for power law inflation.The maximum value $p = 134$
corresponds to $n = 0.985$ and is equivalent to $\sim 67$ e-foldings
of slow-roll inflation.

\bigskip

\leftline {\bf Fig. 3}
\vskip .2 cm
   The biasing factor $b_{16} =  (\dm)^{-1}_{16}$, is shown
plotted against the spectral index $n$, for three values of the Hubble
parameter:
$H = $ 50, 75 \& 100 $ km sec^{-1} Mpc^{-1}$. The dotted lines correspond to
$b_{16} = 1$  and $b_{16} = 2.5 $.

\bigskip

\leftline {\bf Fig. 4}
\vskip .2 cm

  The COBE--normalised spectral energy density
(in units of the critical density)   and the amplitude,$h_{\times+} $,
of gravity waves is shown (in figures $a$ and $b$ respectively) as a
function of the wavelength ( and frequency $f$), for exponential inflation,
quasi-exponential inflation (dotted line), and power law inflation $ a \propto
t^p$
 with $p = 21$ and $p = 9$. For comparison, the expected
sensitivity of the Laser Interferometer Gravitywave Observatory (LIGO),
 (Christensen 1992) and of the projected ``Beam in space" (space -
interferometer)
has also been plotted (Thorne 1988).

\end